\documentclass[a4paper]{jpconf}
\usepackage[latin2]{inputenc}
\usepackage{t1enc}
\usepackage{graphicx} 
\usepackage{cite}
  \usepackage{amssymb,amsmath,epsfig,bm,pifont}
\newcommand{\be}{\begin{equation}}\newcommand{\ee}{\end{equation}}%
\newcommand{\bd}{\begin{displaymath}}\newcommand{\ed}{\end{displaymath}}
\newcommand{\bit}{\begin{itemize}}                                        
 \newcommand{\eit}{\end{itemize}}                                         
\newcommand{\ben}{\begin{enumerate}}                                      
 \newcommand{\een}{\end{enumerate}}                                       
\newcommand{\baa}{\begin{array}{lll}}                                     
 \newcommand{\eaa}{\end{array}}                                           
\newcommand{\ba}{\begin{eqnarray}}                                        
 \newcommand{\ea}{\end{eqnarray}}                                         
\newcommand{\gev}[1]{\relax\ifmmode{\text{GeV}^{#1}}                      
                     \else{GeV$^{#1}${ }}\fi}                             
\def\MSbar{\relax\ifmmode\overline                                        
            {\rm MS}\else{$\overline{\rm MS}${ }}\fi}                     
\def\as{\relax\ifmmode \alpha_s\else{$ \alpha_s${ }}\fi}                  
\def\abar{\relax\ifmmode{\bar{a}}\else{$\bar{a}${ }}\fi}                  
    
\usepackage{color}
\definecolor{mBlue}{rgb}{0,0,1}
 
\definecolor{mRed}{rgb}{1,0,0}
 
\definecolor{mGreen}{rgb}{0.0,1.0,1.0}
 
\begin{document}
\title{Study of Spin through Gluon Poles}

\author{\underline{I.V.~Anikin}$^1$, L.~Szymanowski$^2$, O.V.~Teryaev$^1$
and N.~Volchanskiy$^{1,\,3}$}

\address{$1$ Bogoliubov Laboratory of Theoretical Physics, JINR,
         141980 Dubna, Russia}

\address{$2$ National Centre for Nuclear Research (NCBJ),
            00-999 Warsaw, Poland}

\address{$3$ Research Institute of Physics, Southern Federal University,
             344090 Rostov-on-Don, Russia}

\ead{anikin@theor.jinr.ru}

\begin{abstract}
Based on the use of contour gauge and collinear factorization,
we propose a new set of single spin asymmetry which can be measured in polarized Drell-Yan process by SPD@NICA.
We stress that all of discussed  single spin asymmetries exist owing to the gluon poles manifesting
in the twist-3 or twist-2$\otimes$twist-3 parton distributions related to 
the transverse-polarized Drell-Yan process.
\end{abstract}

\section{Introduction}\label{sec:sec1}
The studies of hadron structure are based on the investigations of both semi-inclusive and exclusive processes which 
can be described by transverse momentum dependent (TMDs) and generalized parton distributions (GPDs).
The duality and matching between different factorization regimes are one of importances for the coherent QCD description of hadron structure.
We discuss the manifestation of such effects in the pion-nucleon Drell-Yan process at large $x_F$ provided
pions are described by wave functions and distribution amplitudes rather than parton distributions.

We calculate the gauge invariant Drell-Yan-like hadron tensors.
In connection with new COMPASS results, we predict the new single spin asymmetry which probes gluon poles together with
chiral-odd and time-odd functions.
The relevant pion production as a particular case of Drell-Yan-like process has been discussed.
For the meson-induced Drell-Yan process, we model an analog of the twist three distribution function,
which is a collinear function in inclusive channel,
by means of two non-collinear distribution amplitudes which are associated with
exclusive channel. This modelling demonstrates the fundamental duality between different factorization regimes.

\section{Drell-Yan-like processes: new result for pion-nucleon Drell-Yan process}\label{sec:sec2}
\begin{figure}[t]
\centerline{\includegraphics[width=0.4\textwidth]{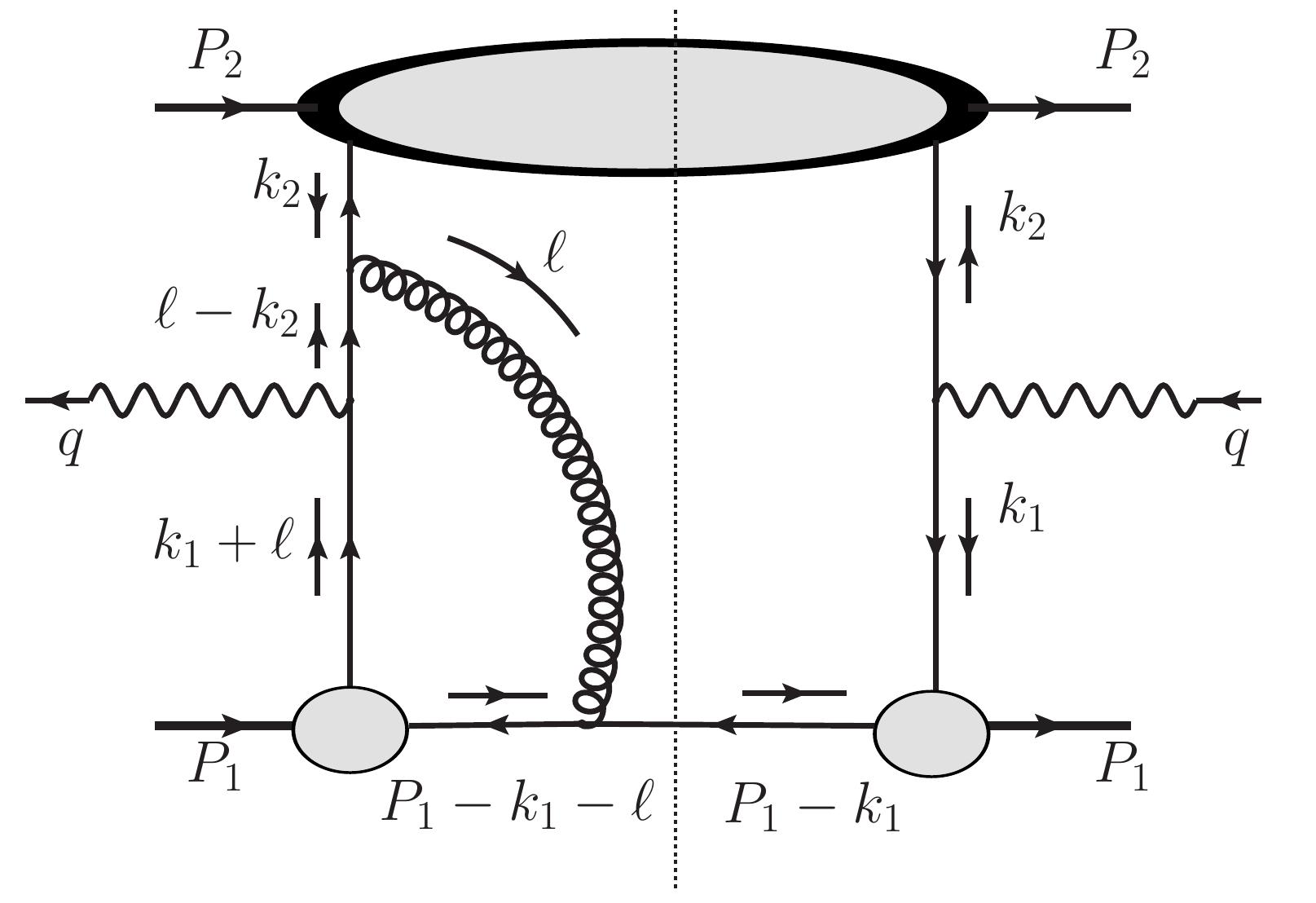}
\hspace{1.cm}\includegraphics[width=0.4\textwidth]{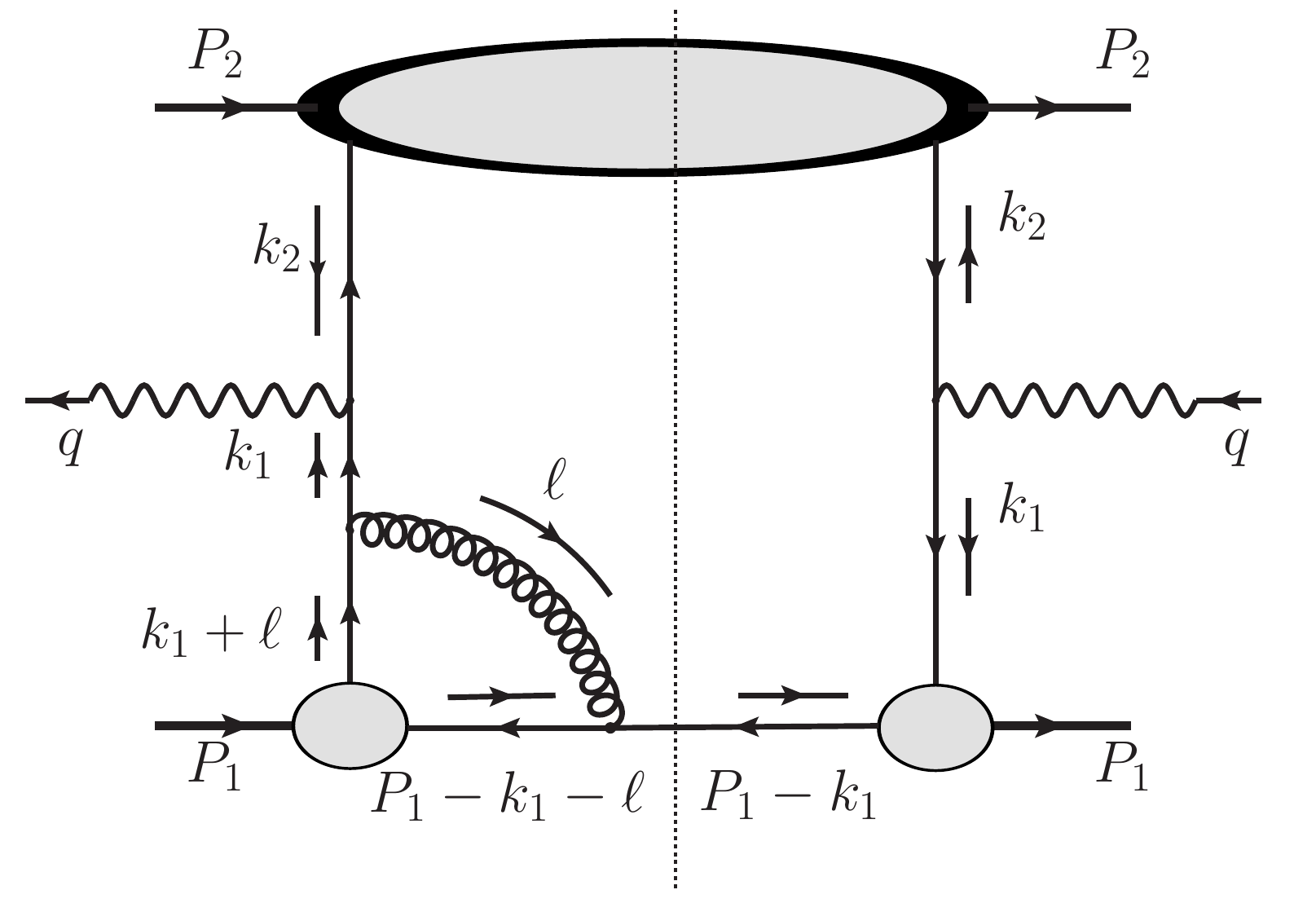}}
\caption{The Feynman diagrams which contribute to the polarized Drell-Yan hadron tensor:
the left panel corresponds to the standard diagram while the right panel -- to the nonstandard diagram.
Note that the standard and nonstandard
diagrams differ by the hard parts \cite{Anikin:2006hm}.
}
\label{Fig-DY}
\end{figure}

We study the process which has the following schematic representation, see Fig.\ref{Fig-DY}
\begin{eqnarray}
A^{(\uparrow\downarrow)}(p_1) + B(p_2) \to \gamma^*(q) + X(P_X)\to
\ell(l_1) + \bar\ell(l_2) + X(P_X),
\end{eqnarray}
where $A$ denotes the (transverse)polarized nucleon, $B$ stands for an arbitrary hadron and
$l_1+l_2=q$ has a large mass squared ($q^2=Q^2$).
The single transverse spin asymmetry (SSA) under our consideration is given by
\begin{eqnarray}
\label{SSA-1}
{\cal A}= \frac{d\sigma^{(\uparrow)} - d\sigma^{(\downarrow)}}{d\sigma^{(\uparrow)} + d\sigma^{(\downarrow)}},
\quad
\frac{d\sigma^{(\uparrow\downarrow)}}{d^4 q d\Omega}=\frac{\alpha^2_{em}}{2j q^4}
{\cal L}_{\mu\nu}\, H_{\mu\nu}\, ,
\end{eqnarray}
where ${\cal L}_{\mu\nu}$ is a lepton tensor, and $H_{\mu\nu}$ - the
QED gauge invariant hadron tensor which corresponds to the difference between direct and mirror channels in the regime of $x_F\to 1$.

In Fig.\ref{Fig-DY}, the upper blob with the light-cone minus dominant direction is given by
the following hadron matrix element:
\begin{eqnarray}
\langle P_2, S_\perp | \big[ \bar\psi\, \sigma^{-\perp} \,
\psi \big]^{\text{tw-2}} |S_\perp, P_2 \rangle\stackrel{{\cal F}}{\sim} \varepsilon^{- \perp S^{\perp} P_{2}}
\, \bar h_{1}(y)
\end{eqnarray}
and the lower blob with the light-cone plus dominant direction corresponds to
\begin{eqnarray}
&&\langle P_1, S | \big[ \bar\psi\,\gamma^+\,
A^{\alpha}_\perp \,\psi \big]^{\text{tw-3}}|S, P_1 \rangle
\Big|_{\text{DY}}
\Longrightarrow
\\
&&
{\cal D}^{\alpha\beta} \, \langle 0 | \big[\bar\psi\gamma^+(\gamma_5)\psi \big]^{\text{tw-2}}| S, P_1 \rangle \, \gamma_\beta \,
 \langle P_1, S | \big[\bar\psi\sigma^{- +}(\gamma_5)\psi\big]^{\text{tw-3}} | 0 \rangle
\Big|_{\pi N\text{-DY}}
\nonumber
\end{eqnarray}
which finally leads to (see, \cite{Anikin:2017azc,Anikin:2016bor,Anikin:2010wz,Anikin:2015xka,Anikin:2015esa} for details)
\begin{eqnarray}
\label{B-fun}
\widetilde{B}(x_1,x_2)=\frac{1}{2}
\frac{\Phi_{(1)}^{\text{tw-3}}(x_1)
\stackrel{\hspace{0.2cm}\vec{\bf k}_{1}^{\,\perp}}{\circledast}
\Phi_{(2)}^{\text{tw-2}}(x_2)}{x_2-x_1  - i\epsilon}\,.
\end{eqnarray}
where $- i\epsilon$ stems from the contour gauge: $[z^-,-\infty^-]_{A^+}=1$.

For the chiral-odd contributions, within our framework, we predict a new SSA which reads
\begin{eqnarray}
\label{NewSSA}
{\cal A}_T=\frac{S_{\perp}}{Q} \, \frac{  D_{[\sin 2\theta]} \, \sin\phi_S B^{\sin\phi_S}_{UT}}
{\bar f_1(y_B)\,H_1(x_B)},\, D_{[\sin 2\theta]}=\frac{\sin 2\theta}{1+\cos^2\theta}
\end{eqnarray}
where
\begin{eqnarray}
B^{\sin\phi_S}_{UT}=2\bar h_1(y_B)\, \Phi_{(1)}^{\text{tw-3}}(x_B)\hspace{-0.2cm} \stackrel{\hspace{0.2cm}\vec{\bf k}_{1}^{\,\perp}}{\circledast}
\hspace{-0.1cm}\Phi_{(2)}^{\text{tw-2}}(x_B)
\end{eqnarray}
and $\bar f_1(y_B),\,H_1(x_B)$ emanate from the unpolarized cross-section and they parameterize the following matrix elements.
The result is presented in terms of \cite{COMPASS}. We want to emphasize that the predicted SSA (\ref{NewSSA}) exists due to 
the complexity of $ \widetilde{B}(x_1,x_2)$-function (\ref{B-fun}), {\it i.e.} $\Im\text{m} \widetilde{B}(x_1,x_2)\neq 0$.

From comparison with recent COMPASS results on pion-nucleon Drell-Yan process, we can see that our ${\cal A}_T$ is completely new one.  
Indeed, the leading twist Sivers asymmetry $A^{\sin\phi_S}_{UT}$
which formally stands at the similar tensor combination
$\varepsilon^{l^\perp_1 S^\perp P_1 P_2}$, appears only together with the depolarization factor $D_{[1+\cos^2\theta]}$.
In its turn, the higher twist asymmetries $A^{\sin(\phi_S\pm\phi)}_{UT}$ at $D_{[\sin 2\theta]}$
correspond to the different tensor structures, $\varepsilon^{q S^\perp P_1^\perp P_2}\sim \sin(\phi_S\pm\phi)$. 

\section{Factorization theorem, in a nutshell}\label{sec:sec3}
Since the main our tool is the factorization procedure, let us outline the principal items of factorization we adhere. 
With a help of factorization, a given amplitude of hadron tensor of any hard processes can be estimated, not calculated,
by the well-defined procedure within the corresponding asymptotical regime. As a result of this estimation, we arrive at the 
well-known factorization theorem which states that 
the short (hard) and long (soft) distance
dynamics can be separated out provided $Q^2$ is large, {\it i.e.}
\begin{eqnarray}
&&\hspace{-0.5cm}{\cal W}_{\mu\nu...}=
\int d^4 k_1 d^4 k_2 ... \,\text{tr} \big[\Phi_1 (k_1) E_{\mu\nu...}(k_1,k_2,...) \Phi_2 (k_2) ... \big]
\stackrel{Q^2\to\infty}{ \Longrightarrow}
\nonumber\\
&&\int dxdy ... \,\text{tr} \big[\Phi_1 (y) E_{\mu\nu}(x,y,...)\,\Phi_2 (x)... \big] +
{\cal O}(1/Q^2)
\end{eqnarray}
where $E_{\mu\nu...}(k_i)$ implies the products of propagators multiplying by the momentum conservation $\delta$-function, and
\begin{eqnarray}
&&
E_{\mu\nu}(x,y,...| {\bf k}^T_i=0),\quad
\Phi(k)\stackrel{{\cal F}}{=} \langle {\cal O}^{(\bar\psi, \psi, A)}(z,0) \rangle\,,
\\
\label{PDF-1}
&&
\Phi(x_i) = \int dk^+_i \, \delta(x_i-k^+_i\cdot n) \,\int dk^-_i\, d^2{\bf k}^T_i \, \Phi(k_i)\,.
\end{eqnarray}
Besides, both the hard and soft parts should be independent of each other,
UV- and IR-renormalizable and, finally, parton distributions must possess
the universality property.
We notice that the transverse momentum integral in (\ref{PDF-1}) gives us the possibility 
for the study of TMDs. In the other words, the TMDs can contribute through the corresponding moments. 
However, if the nontrivial ${\bf k}^T_i$-dependence has been kept in $E_{\mu\nu}(x,y,...| {\bf k}^T_i)$,
the factorization theorem can be in question. It can be happened, for example, in case we want to 
restore the ${\bf k}^T$-dependent Wilson line in the correlators $\Phi(k_i)$. 

\section{Contour gauge and complexity of $B(x_1,x_2)$-functions}\label{sec:sec4}
We show that the contour gauges for gluon fields play
the crucial role for our study. The contour gauge belongs to the class of non-local  gauges
that depends on the path connecting two points in the correlators.
It turns out, in the cases which we consider, the prescriptions for the gluonic poles in the twist $3$ correlators,
see (\ref{B-fun}), are dictated by the contour gauge $[x,\,\pm\infty]=1$. Indeed, 
one can easily check that
the representation with $x_1-x_2 + i\epsilon$ in the denominator belongs to the gauge $[x,\,-\infty]=1$, while
the representation  with $x_1-x_2 - i\epsilon$ belongs to the gauge $[+\infty,\, x]=1$.

Thus, we find that the nonstandard new terms,
which exist in the case of the complex twist-$3$ $B$-function with the corresponding prescriptions,
do contribute to the hadron tensor exactly as the standard term known previously.
This is  another important result of our work.

In conclusion of this section, let us briefly discuss the main items of the contour gauge conception.
To describe this class of gauges, we adhere the geometrical interpretation of gluons
where the gluon field is a connection of the principal fiber bundle
${\cal P}(\mathbb{R}^4, G, \pi)$.
Here, $\mathbb{R}^4$ is the base where the principal fiber bundle is determined, $G$ denotes the group
defined on the given fiber and $\pi$ is a transformation of the base $\mathbb{R}^4$ into the fiber bundle
${\cal P}$). The corresponding path-dependent Wilson factor (or line), representing each element $\textbf{g}(x)$ of the fiber, 
defines the gauge-transformed field:
\begin{eqnarray}
\label{Ag}
A^{\textbf{g}}_\mu(x)=\textbf{g}^{-1}(x)A_{\mu}(x) \textbf{g}(x) +
\frac{i}{g}\textbf{g}^{-1}(x)\partial_{\mu}\textbf{g}(x).
\end{eqnarray}
The set of these fields for all $\textbf{g}(x)$ forms the gauge orbit.
In order to quantize a system of the gauge fields, one chooses the only element of each
orbit. In contrast to the standard way, we first fix an arbitrary
point $(x_0, \textbf{g}(x_0))$ in the fiber. Then, we define two directions: one of them in the base, the other in
the fiber. The direction in the base $\mathbb{R}^4$ is the tangent vector of a curve which
goes through the given point $x_0$. At the same time, the direction in the fiber can be uniquely determined as the
tangent subspace related to the parallel transition. Due to this procedure, we uniquely
define the point in the fiber bundle.

Having solved the parallel transport equation defined on the fiber as
\begin{eqnarray}
\frac{dx_\alpha(v)}{dv} \, {\cal D}_\alpha \textbf{g}(x(v)) = 0\,,
\end{eqnarray}
we find the solution in terms of the Wilson line:
\begin{eqnarray}
\label{gg}
\textbf{g}(x)=Pexp\Big\{ ig \int\limits_{\mathbb{P}(x_0,x)} d\omega\cdot A(\omega)\Big\} \textbf{g}(x_0) ,
\end{eqnarray}
where the points $x_0$ and $x$ are connected by the path $\mathbb{P}$.
The starting point $x_0$ is usually fixed. Here, $\textbf{g}(x_0)$ is chosen to be equal to unity.
Note that the fixing of $\textbf{g}(x)$ ensures a unique choice of the element in the orbit.
We now insert (\ref{gg}) into (\ref{Ag}) and can see that the field $A^{\textbf{g}}_\mu(x)$ is completely
determined by the form of the path which connects the starting and final points. Moreover, using (\ref{Ag}) and
(\ref{gg}), one obtains the property:
\begin{eqnarray}
\label{propAg}
Pexp\Big\{ ig \int\limits_{x_0}^{x} d\omega\cdot A^{\textbf{g}}(\omega)\Big\} =
\textbf{g}^{-1}(x) Pexp\Big\{ ig \int\limits_{x_0}^{x} d\omega\cdot A(\omega)\Big\}\textbf{g}(x_0) .
\end{eqnarray}
Then, if we insert (\ref{gg}) into (\ref{Ag}),
we arrive at
\begin{eqnarray}
\label{cg1}
A^{\textbf{g}}_\mu(x)=\int\limits_{\mathbb{P}(x_0,x)} dz_\alpha \frac{\partial z_\beta}{\partial x_\mu}\,
\textbf{g}^{-1}(z) \,G_{\alpha\beta}(z| A)\, \textbf{g}(z)=
\int\limits_{\mathbb{P}(x_0,x)} dz_\alpha \frac{\partial z_\beta}{\partial x_\mu}\,
G_{\alpha\beta}(z| A^{\textbf{g}}) ,
\end{eqnarray}
where
\begin{eqnarray}
\label{Gcontour}
G_{\alpha\beta}(z| A^{\textbf{g}})\equiv G_{\alpha\beta}^{\textbf{g}}(z)
=\textbf{g}^{-1}(z) \,G_{\alpha\beta}(z| A)\, \textbf{g}(z)
\end{eqnarray}
The contour gauge condition demands that $\textbf{g}(x)$ is equal to
unity for all $x$ belonging to the base, {\it i.e.}
\begin{eqnarray}
\label{cg2}
[x,\,x_0]\stackrel{def}{=}Pexp\Big\{ ig \int\limits_{x_0}^{x} d\omega\cdot A(\omega)\Big\} = 1 ,
\,\,\, \forall x \in \mathbb{R}^4 .
\end{eqnarray}
Therefore, within the contour gauge, the field $A^{\textbf{g}}_\mu$ (see, (\ref{cg1})) becomes
\begin{eqnarray}
\label{cg3}
A^{c.g.}_\mu(x)=
\int\limits_{\mathbb{P}(x_0,x)} dz_\alpha \frac{\partial z_\beta}{\partial x_\mu}\,
G_{\alpha\beta}(z| A^{c.g.}) ,
\end{eqnarray}
{\it i.e.} the gluon field $A^{\textbf{g}}_\mu$ is a linear functional of the tensor $G_{\mu\nu}$.

\section{Conclusions}\label{sec:concl}
To conclude we derive the gauge invariant meson-induced DY hadron tensor with the essential
spin transversity and primordial transverse momenta. Our calculation includes both the
standard-like, which is well-known, and nonstandard-like diagram, which is first found in \cite{Anikin:2010wz}, contributions.
The latter contribution plays a crucial role for the gauge invariance.

We study the case of pion distribution amplitudes which has been projected onto the chiral-odd combination.
The latter singles out the chiral-odd parton distribution inside nucleons.
The chiral-odd tensor combinations are very relevant for the future experiments implemented by COMPASS \cite{COMPASS}.
We predict a new single transverse spin asymmetry which can be measured experimentally, for example, by SPD@NICA.
The latter asymmetry can eventually be treated as a signal for the transverse ``primordial'' momentum dependence
which probes both gluon poles and time-odd functions.

\ack
We thank A.V.~Efremov, L.~Motyka and A.~P.~Nagaytsev for useful discussions.
The work by I.V.A. was partially supported by the Bogoliubov-Infeld Program
and Heisenberg-Landau Program HL-2017. L.Sz. is partly supported by 
grant No $2015/17/\text{B}/\text{ST}2/01838$ from the National Science
Center in Poland.

\end{document}